# Chemistry and High Temperature Superconductivity

**J. Paul Attfield**

*Centre for Science at Extreme Conditions and School of Chemistry, University of Edinburgh, Mayfield Road, Edinburgh EH9 3JZ, UK. Fax: 44 131 6517049; Tel: 44 131 6517229; E-mail: j.p.attfield@ed.ac.uk*

## Abstract

Seven distinct families of superconductors with critical temperatures at ambient pressure that equal or surpass the historic 23 K limit for $Nb_3Ge$ have been discovered in the last 25 years. Each family is reviewed briefly and their common chemical features are discussed. High temperature superconductors are distinguished by having a high ($\geq 50\%$) content of nonmetallic elements and fall into two broad classes. 'Metal-nonmetal' superconductors require a specific combination of elements such as Cu-O and Fe-As which give rise to the highest known $T_c$'s, probably through a magnetic pairing mechanism. 'Nonmetal-bonded' materials contain covalently-bonded nonmetal anion networks and are BCS-like superconductors. Fitting an extreme value function to the distribution of $T_c$ values for the known high-$T_c$ families suggests that the probability of a newly discovered superconductor family having maximum $T_c > 100$ K is ~0.1-1%, decreasing to ~0.02-0.2% for room temperature superconductivity.

## 1. Introduction

Superconductors have zero electrical resistance and behave as perfect diamagnets (known as the Meissner effect). This arises from the condensation of electrons near the Fermi level into Cooper pairs that behave as a collective quantum mechanical state – a superconductor is a charged superfluid. Thermal pair-breaking limits superconductivity to a maximum critical temperature ($T_c$) above which the material shows metallic or semiconducting behaviour with a finite resistance. Superconductivity is also limited by critical magnetic fields ($H_c$) and current densities ($J_c$) at temperatures below $T_c$. The critical temperature is mainly determined by chemical composition and structure, whereas the critical fields and currents are also strongly influenced by microstructure and are often not optimum in homogenous materials. The electron-pairing interactions are relatively weak, e.g. in comparison to magnetic exchange interactions, so that all known superconductors have $T_c$'s well below ambient temperature. Cryogenic cooling is thus needed to exploit the useful properties of superconductors such as in power transmission cables, magnetically levitated trains, or SQUID (superconducting quantum interference device) electronics based on the Josephson effect of tunnelling Cooper pairs. Increasing $T_c$ towards ambient temperature is thus a major ambition for superconductivity research.

The first century of superconducting materials research divides into two eras, the 'low-$T_c$' period from the 1911 discovery of the zero-electrical resistance transition in mercury until 1986 when the record critical temperature was $T_c = 23$ K in Nb$_3$Ge, and the subsequent 'high-$T_c$' era, during which several types of chemically complex solids with $T_c$'s up to 138 K have emerged. This review will briefly describe the seven distinct families of high-$T_c$ superconductors (taken to be those with $T_c \geq 23$ K at ambient pressure) that have been discovered in the last quarter century. The families vary in size from containing one to many chemically and structurally similar materials with a common physical mechanism for superconductivity. The overall chemical trends and possible future directions for materials discovery are also discussed.

## 2. High-$T_c$ families

### 2.1 Cuprates

High (critical) temperature superconductivity was born from the discovery of an unprecedented $T_c = 35$ K transition in Ba-doped La$_2$CuO$_4$ in 1986.[1] A flurry of activity in the following years led to the identification of many more superconducting cuprates, with the highest $T_c$ of 138 K found in HgBa$_2$Ca$_2$Cu$_3$O$_{8+\delta}$. A fluorinated sample of this phase showed an onset $T_c$ of 166 ± 1 K at 23 GPa pressure which is the highest measured superconducting critical temperature to date,[2] and approaches the lowest recorded terrestrial temperature of 184 K. An enormous literature is available for the cuprate superconductors; some recent books and review articles are cited here.[3,4,5,6,7] Although the pairing mechanism and a convincing explanation for the magnitude of $T_c$ in this family remain controversial, the essential chemical features are clearly established.

The complex chemistry of the cuprates results from the requirement for several structural features, shown schematically in Fig. 1(a), to optimise superconductivity;

1  Copper oxide planes are essential. These have stoichiometry CuO$_2$ and a geometry like that found in the MO$_2$ planes of the AMO$_3$ perovskite structure (Fig. 1(b)). Hence cuprates are sometimes described as 'layered perovskites'. Maximum $T_c$'s are generally found for materials with blocks of three adjacent, hole-doped CuO$_2$ planes.

2  Electropositive cations, usually from the alkaline earth (Ca, Sr, Ba) or rare earth (La-Lu, Y) metals, act as layer separators in two distinctive structural roles. Large 'A' cations (typically Sr$^{2+}$, Ba$^{2+}$ or La$^{3+}$) support additional coordination of a further oxygen to copper, and this can provide a connection to additional metal (M) oxide layers. Smaller electropositive 'B' cations (usually $Ca^{2+}$ or a small rare earth) separate $CuO_2$ planes in multilayer cuprates without allowing intercalation of O between Cu's in adjacent planes which is detrimental to superconductivity. n $CuO_2$ planes require (n-1) B cation spacer layers.

3 Blocks of one or two covalent metal oxide layers $MO_x$ (M can be Cu, Ru, Hg, Tl, Pb, Bi) layers are connected to $CuO_2$ planes via interplanar oxides in the AO layers. The $MO_x$ layers are sometimes termed the 'charge reservoir' as they compensate for the doping of the $CuO_2$ planes although this can also achieved by non-aliovalent substitutions at the A or B sites.

Hence many cuprates have compositions $(MO_x)_m(AO)_2B_{n-1}(CuO_2)_n = M_mA_2B_{n-1}Cu_nO_z$, often abbreviated as M-m2(n-1)n, e.g. the highest-$T_c$ material $HgBa_2Ca_2Cu_3O_{8+\delta}$ is abbreviated as Hg-1223.

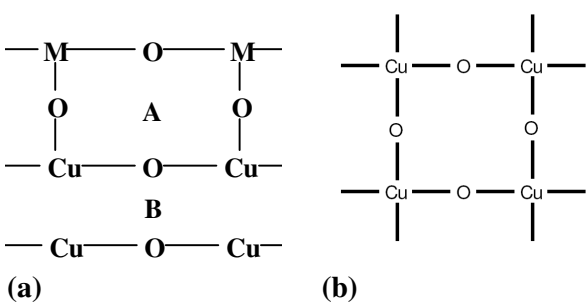

**(a)**          **(b)**

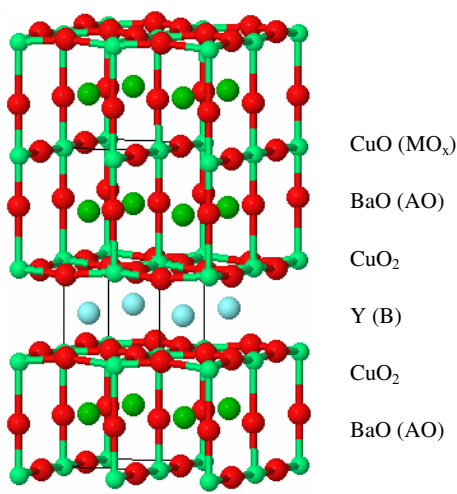

CuO ($MO_x$)

BaO (AO)

$CuO_2$

Y (B)

$CuO_2$

BaO (AO)

**(c)**

**Fig. 1** Structural features of the high-$T_c$ cuprate superconductors (a) schematic view of the key structural components, (b) a $CuO_2$ plane, (c) the crystal structure of $YBa_2Cu_3O_7$ with the repeat sequence of layers labelled following (a).

The undoped cuprates contain $Cu^{2+}$ within the $CuO_2$ sheets and are antiferromagnetic insulators, with very strong antiferromagnetic Cu-O-Cu superexchange interactions of coupling strength J/k ~ 1500 K. Superconductivity is induced by doping the $CuO_2$ sheets; electron doping (reduction of $Cu^{2+}$) is effective in a few cases, e.g. $Nd_2CuO_4$, but for the majority including all the highest-$T_c$ materials, hole-doping is achieved by cation substitutions, increasing oxygen content, or by band overlap. The latter two mechanisms are found in many materials, e.g. maximum $T_c$'s ≈ 40 K are obtained for x = 0.16 and y = 0.08 in the $La_{2-x}Sr_xCuO_4$ and $La_2CuO_{4+y}$ systems respectively, both of which yield an average +2.16 oxidation state for Cu. Hole-doping through band overlap is less common, but is well-established in the important 93 K superconductor $YBa_2Cu_3O_7$ (also known as YBCO or (Y)123). This contains distinct CuO chains as the $MO_x$ layers in the above structural classification (see Fig. 1(c)), and Cu band overlap results in a formal charge distribution $Cu^{2.6+}Ba_2YCu^{2.2+}_2O_8$. Similar Ru-Cu charge transfer induces superconductivity in the ruthenocuprate $Ru^{4.8+}Sr_2GdCu^{2.1+}_2O_8$.[8]

A simplified electronic phase diagram for the hole-doped cuprates is shown in Fig. 2. Initial doping of a $Cu^{2+}$ parent material disrupts long range antiferromagnetic order and the Néel transition is suppressed at 3% doping. Superconductivity appears above 5% doping – the intermediate region is found to be physically inhomogenous with both magnetic and superconducting correlations present. Further oxidation increases $T_c$ to a maximum at 15-20% doping, above which superconductivity is suppressed and is no longer apparent above ~25% hole doping. There is substantial evidence for a diffuse high temperature insulator-metal transition at which a pseudo-gap opens,[3] and this coincides with $T_c$ for the overdoped superconductors. Above $T_c$, the cuprates have unusual normal state electronic properties that evidence strong electron-electron correlations, but at high doping levels they become more like conventional metals.

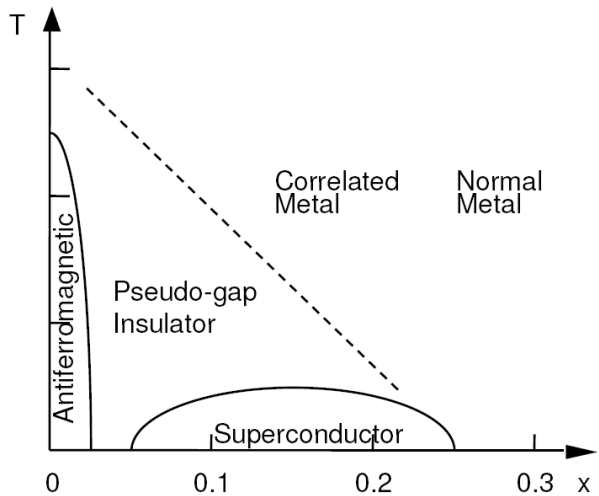

**Fig. 2** Schematic electronic phase diagram for cuprate superconductors as a function of the hole doping x (equivalent to average Cu oxidation state 2+x).

In addition to the doping level, structural features are also very important for optimising $T_c$ in the cuprates. Buckling of the Cu-O-Cu bridges in the $CuO_2$ sheets suppresses superconductivity and favours alternative charge and spin ordered insulating ground states, so large $Ba^{2+}$ cations at the A type cation sites help to preserve flat $CuO_2$ planes and high $T_c$'s. Disorder arising from mixed A or B cations adjacent to the planes also suppresses superconductivity. Hence the highest reported $T_c$ for single layer cuprates is 98 K for optimally doped $HgBa_2CuO_{4+\delta}$ ($\delta$ = 0.08) which has only $Ba^{2+}$ cations adjacent to the $CuO_2$ planes, and relatively little additional strain and disorder from the small concentration $\delta$ of oxygen interstitials between the Hg sites.

Coupling between nearby $CuO_2$ layers separated by B cation layers also enhances $T_c$; the highest values are found in the Hg-family where $T_c$ increases up to 138 K for n = 3. $T_c$ decreases for higher n most probably because the doping becomes non-uniform across inequivalent $CuO_2$ planes.

The pairing mechanism for superconductivity in the cuprates remains unclear. The essential features for theoretical descriptions are the d-wave symmetry of the order parameter (the wavefunction describing the Cooper pairs has the symmetry of a $d_{x^2-y^2}$ atomic orbital), the presence of strong antiferromagnetic correlations, and the pseudogap feature. A plausible explanation is that antiferromagnetic fluctuations mediate the pairing instead of the electron-phonon coupling found in conventional BCS (Bardeen, Cooper and Schrieffer) type materials.

### 2.2 Fullerides

The discovery of high temperature superconductivity in copper oxides was followed by another remarkable finding from a very different group of materials. After buckminsterfullerene ($C_{60}$) was first identified and isolated in the 1980's, alkali metal fulleride derivatives were synthesised and superconductivity was first reported in $K_3C_{60}$ with $T_c$ = 19 K.[9] The fulleride superconductors are highly air-sensitive which hinders their characterisation and limits their practical utility. Nevertheless, the variation of superconductivity across the $A_3C_{60}$ family has been explored in detail and related alkaline earth and lanthanide-doped fullerides, with $T_c$'s up to 8 K for $Ca_5C_{60}$ and $Sm_{2.75}C_{60}$, have also been prepared.[10,11,12]

The main tuning parameter for the $A_3C_{60}$ superconductors is the interfulleride separation, represented by the unit cell volume per $C_{60}$, as shown in Fig. 3. $T_c$ increases up to a maximum of 33 K for $RbCs_2C_{60}$ as the volume increases, but beyond this limit superconductivity is destabilised with respect to an antiferromagnetic magnetic ground state. Recent studies have found that $T_c$ increases under pressure up to 35 and 38 K in the face- and body- centred cubic polymorphs of $Cs_3C_{60}$, respectively.[13]

The phonon-mediated BCS mechanism for superconductivity in the weak coupling limit describes many aspects of the fullerides. Vibrational spectra reveal a coupling between the conduction electrons and high frequency vibrations of the $C_{60}^{3-}$ anions that strengthens as the interfulleride separation increases, and a BCS-type [13]C isotope effect is also observed. However, the transition to a Mott (magnetic) insulating state at limiting high separations is more similar to the breakdown of metallic and superconducting behaviour at low dopings in cuprates and other unconventional high-$T_c$ superconductors.

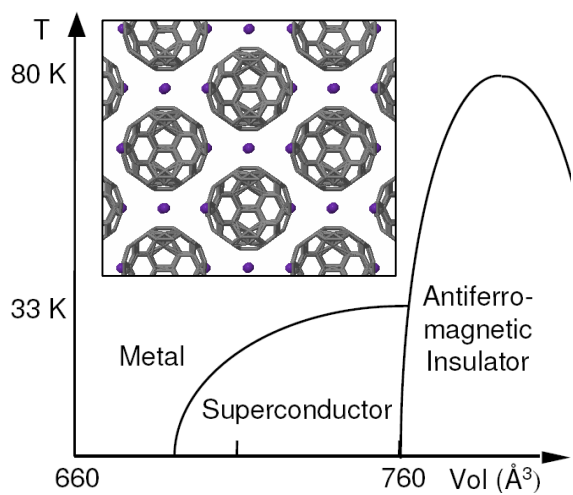

**Fig. 3** Schematic phase diagram for the $A_3C_{60}$ fulleride superconductors as a function of the volume per $C_{60}$ unit (adapted from ref. 12). The inset shows the face-centred cubic $A_3C_{60}$ structure.

## 2.3 Barium bismuthate

The perovskite $BaBiO_3$ undergoes charge disproportionation that results in a distorted crystal structure containing an ordered alternation of $Bi^{3+}$ and $Bi^{5+}$ sites (Fig. 4). Suppression of charge order in $Ba_2Bi^{3+}Bi^{5+}O_6$ to give a superconductor was first demonstrated by substituting Pb for Bi, and the maximum $T_c$ in this system is 13 K.[14] Renewed interest following the discovery of superconductivity in cuprates led to the further discovery of $T_c$'s up to 30 K in $Ba_{1-x}K_xBiO_3$.[15] The phase diagram for this system[16] in Fig. 4 shows that superconductivity emerges with the maximum $T_c$ immediately beyond the suppression of the charge ordered state at x = 0.38 and diminishes with further doping up to the x = 0.5 stability limit of the solid solution.

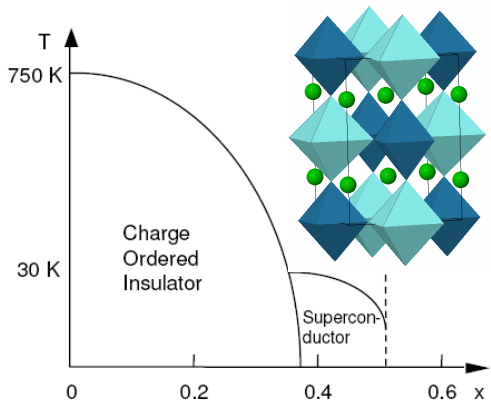

**Fig. 4** Schematic $Ba_{1-x}K_xBiO_3$ phase diagram (temperature not to scale) showing the onset of superconductivity above the x = 0.38 limit of the charge ordered phase. The inset shows the charge ordered perovskite superstructure of $Ba_2Bi^{3+}Bi^{5+}O_6$ with light/dark shading of the $Bi^{3+}/Bi^5$ octahedra.

As superconductivity is only found in the three-dimensional bismuthate perovskite structure there are few chemical variations. $Sr_{1-x}K_xBiO_3$ and $Sr_{1-x}Rb_xBiO_3$ analogues have been prepared at high pressure, but these have lower $T_c$'s of 12 and 13 K respectively.[17] The large radii and good size matching of $Ba^{2+}$ and $K^+$ ions appear to be optimal for superconductivity in doped-$BaBiO_3$.

Optical spectra show that $Ba_{1-x}K_xBiO_3$ is an s-wave superconductor (like conventional low-$T_c$ materials),[18] but the electron-phonon coupling constant was found to be too small for conventional BCS coupling to explain the high $T_c$. Electron-phonon coupling may be enhanced by electron-electron interactions in negative-(Hubbard) U models, which follow the negative-U description of the disproportionation of the average $s^1$ configuration to give ordered $s^0$ and $s^2$ states in the BaBiO$_3$ parent material.

## 2.4 Quaternary borocarbides

Many rare earth transition metal borides and carbides are superconducting but most have low T$_c$ values. However a specific family of quaternary materials with composition RM$_2$B$_2$C (R = rare earth, M = Ni, Pd) were discovered to have T$_c$'s up to the past record value of 23 K for YPd$_2$B$_2$C.[19] The structure (Fig. 5) consists of layers of isolated linear B$_2$C groups between R and M layers. Magnetic R cations suppress superconductivity completely for R = Pr, Nd, Sm, Gd, and Tb in the RNi$_2$B$_2$C series and a variety of antiferromagnetic states is found, but coexistence of superconductivity and magnetic order is observed for the later R = Dy, Ho, Er, Tm.[20,21,22] This strong coupling of the rare earth metal moments to the conduction electrons is in contrast to the RBa$_2$Cu$_3$O$_7$ and RFeAsO series where the R-magnetism has little influence on superconductivity. The RM$_2$B$_2$C superconductors appear to be s-wave materials but with an anisotropic energy gap.

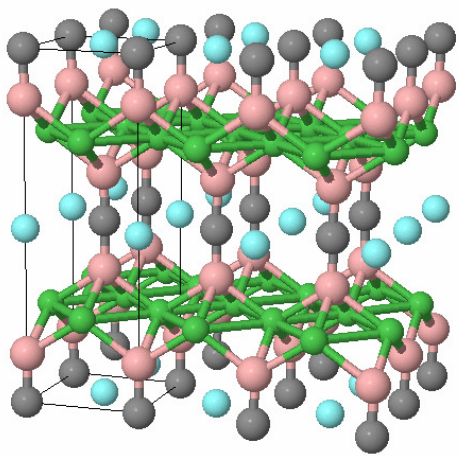

**Fig. 5** Crystal structure of the YPd$_2$B$_2$C superconductor showing metal-metal bonded Pd layers connected by linear BCB units, with Y atoms in the C plane.

## 2.5 Intercalated Nitride Halides

MNX (M = Zr, Hf; X = Cl, Br, I) phases are insulators that contain hexagonal X(MN)$_2$X layers which may be stacked in several polymorphic arrangements. Chemical or electrochemical intercalation of alkali metals (Li, Na, K) into the van der Waals gaps between the layers (Fig. 6(a)), or removal of a small amount of halogen X, dopes electrons into the M d-band inducing superconductivity.[23,24] The maximum observed T$_c$ in this family is in the original report of 25.5

K for $Li_{0.48}(THF)_yHfNCl$ containing cointercalated tetrahydrofuran solvent (THF).[25] Intercalation staging has been observed in $Na_xHfNCl$, and $T_c$'s of 24 and 20 K were reported for the stage 1 and 2 materials respectively.[26] The intercalated MNX phases are very air-sensitive which has hampered study of their physical properties.

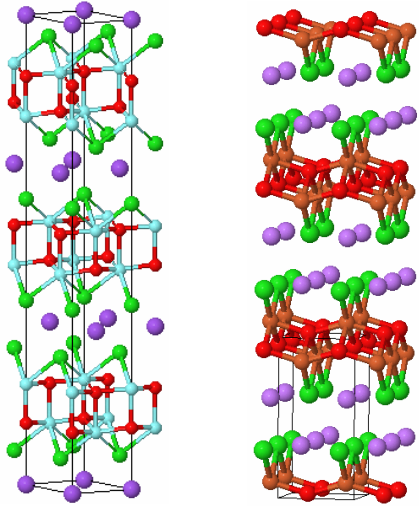

**(a)**                    **(b)**

**Fig. 6** Crystal structures of (a) $Na_xHfNCl$ and (b) $Li_xTiNCl$ showing alkali metals intercalated between hexagonal and orthorhombic MNCl layers respectively.

Superconductivity with maximum $T_c$ = 16.3 K has recently been reported in $A_xTiNCl$ (A = Li, Na, K, Rb),[27] where the constituent layers have an orthorhombic (FeOCl type) structure that differs from that of the Zr and Hf based materials (Fig. 6(b)). This reveals that superconductivity is not specific to one underlying lattice symmetry in this family. Physical measurements show that the nitride halides are not conventional BCS superconductors and have a large superconducting gap ratio $2\Delta/k_BT_c$ = 4.6–5.6. In the $Li_xZrNCl$ system, the maximum $T_c$ is observed for minimum x = 0.06 doping, below which a magnetic insulating state is observed.[28]

## 2.6 Magnesium diboride

The discovery of high-$T_c$ superconductivity at 39 K in $MgB_2$ in 2001 was one of the most unexpected developments in this field.[29] Transition metal diborides were explored extensively during the low-$T_c$ era but the magnesium analogue, which is a standard, air-stable chemical reagent, was thought to be uninteresting as it is has no available d-states. Subsequent measurements have shown that MgB$_2$ can have a T$_c$ of up to 41 K in thin films and can be processed into practical conductors, with critical current densities of up to J$_c$ = 3.4 × 10$^7$ Acm$^{-2}$ reported.[30]

MgB$_2$ has a layered structure with magnesium atoms between graphitic boron sheets (Fig. 7). It appears to be an optimum superconductor 'as is', and doping or substitutions of other metals for Mg or of C for B have not enhanced T$_c$. However, this discovery has inspired reinvestigation of graphite intercalation compounds resulting in several new superconductors with T$_c$'s up to 11.5 K in CaC$_6$.[31] Extensive physical measurements have shown that MgB$_2$ is a near-perfect BCS superconductor, with an isotope-effect increase of approximately 1 K in T$_c$ when $^{11}$B is replaced by $^{10}$B.[32] An important feature is the presence of two superconducting gaps with 2Δ/k$_B$T$_c$ values of 1.1 and 4.0 resulting from Cooper pairing of electrons in σ and π-bands, respectively. The latter value is close to the BCS s-wave value of 3.53. Coupling between the two gap pairings results in the single, observed superconducting transition.

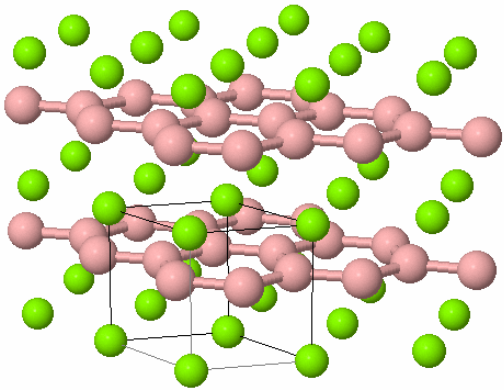

**Fig. 7** The hexagonal structure of the MgB$_2$ superconductor showing graphitic boron sheets interleaved by layers of magnesium.

### 2.7 Iron Arsenides

The final family provides an appropriately symmetric closure to the first quarter century of high temperature superconductor discovery, with very high T$_c$'s second only to those of the cuprates, and many chemical and physical similarities.[33] They are based on FeAs layers in which Fe is tetrahedrally coordinated by As atoms (Fig. 8) and several structure types with different additional layers are known. A comprehensive review of the field has recently been published.[34]

High-T$_c$'s were first reported in the electron-doped LaFeAsO$_{1-x}$F$_x$ series[35] and subsequent study of rare earth RFeAsO$_{1-x}$F$_x$ and oxygen-deficient RFeAsO$_{1-δ}$ analogues led to discovery of the highest T$_c$ = 55 K to date in SmFeAsO$_{1-x}$F$_x$.[36] Hole-doped materials are also superconducting; La$_{0.85}$Sr$_{0.15}$FeAsO has T$_c$ = 25 K;[37] and in the related AFe$_2$As$_2$ and AFeAs families, Ba$_{0.6}$K$_{0.4}$Fe$_2$As$_2$ has T$_c$ = 38 K [38] and LiFeAs has T$_c$ = 18 K.[39]

Superconductivity is also observed in analogues where Fe or As are replaced by similar elements, but with lower T$_c$'s. The binary phase Fe$_{1+x}$Se containing only the FeAs-type layers is superconducting with T$_c$ = 9 K for a small Fe excess x = 0.01,[40] and T$_c$ increases up to 14 K for Fe$_{1+x}$Se$_{0.6}$Te$_{0.4}$. LaFePO has T$_c$ = 6.6 K.[41] Non-Fe analogues such as LaNiPO and LaNiAsO, BaM$_2$As$_2$ (M = Cr, Mn, Co, Ni, Ru, Rh) and BaM$_2$P$_2$ (M = Ni, Rh, Ir), and LiFeP have T$_c$'s <5 K.

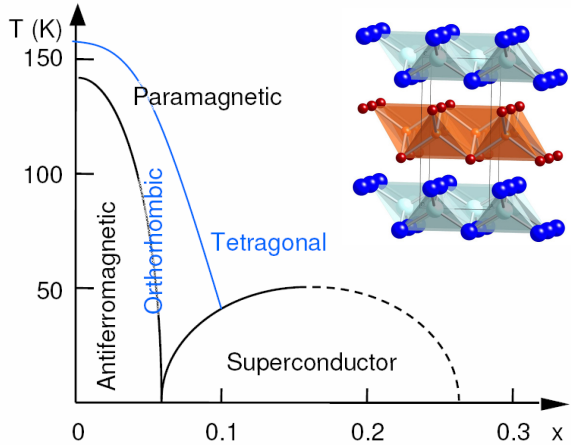

**Fig. 8** Schematic electronic and structural phase diagram for RFeAsO$_{1-x}$F$_x$ iron arsenide superconductors as a function of the electron doping x (equivalent to average Fe oxidation state 2-x). The inset shows the stacking of RO and FeAs slabs in the crystal structure.

The electronic phase diagram for the high-T$_c$ iron arsenide superconductors (Fig. 8) appears similar to that of the cuprates, although an important difference is that the parent materials are metallic whereas the undoped cuprates are Mott insulators. Hence the observed antiferromagnetic order of small (0.1-1 $\mu_B$) Fe moments is probably a spin density wave rather than an array of local moments. Doping suppresses the long range antiferromagnetism and superconductivity emerges with a high T$_c$ for 10-20% doping. The competing ground states stabilise different lattice symmetries. The spin order is antiferromagnetic along one of the two in-plane axes but ferromagnetic in the perpendicular direction, which drives an orthorhombic distortion of the ideal tetragonal structure. However, superconducting phases are tetragonal as the spin density wave is suppressed. The structural transition occurs above the Néel temperature, as shown in Fig. 8, and probably reflects the onset of antiferromagnetic fluctuations. Coexisting antiferromagnetic and superconducting phases are found in the crossover region for some systems. T$_c$ is very sensitive to lattice effects and is optimised when the As-Fe-As angle in the tetrahedral layers is close to the geometric ideal of 109.5°,[42] as shown in Fig. 9.

Two gaps with s-wave symmetry have been found in many experiments and the values of $2\Delta/k_B T_c$ fall into the range 1-4 and 5-9. As for $MgB_2$, coupling between the two orders results in a single superconducting transition. The emergence of superconductivity from a consistent magnetic state in the iron arsenides provides strong evidence for an antiferromagnetic spin fluctuation pairing model, as for the cuprates.

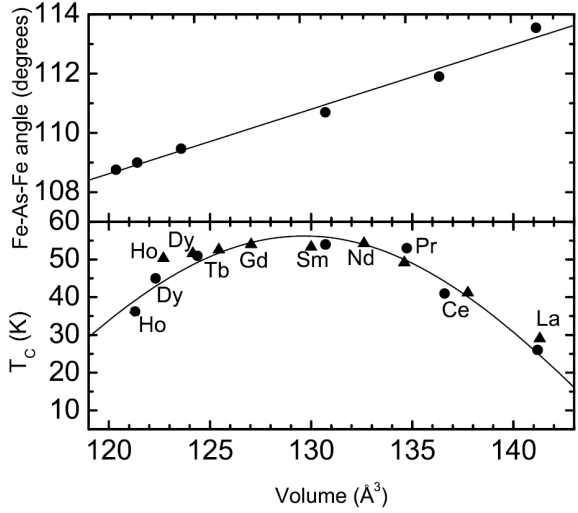

**Fig. 9** Variations of Fe-As-Fe angle (upper panel) and the maximum reported $T_c$ (lower panel) with unit cell volume for optimally doped $RFeAsO_{1-x}F_x$ (circles) and $RFeAsO_{1-\delta}$ (triangles) superconductors, with rare earths R as shown. The overall maximum $T_c = 56$ K obtains for an angle of 110.6°, close to the ideal tetrahedral angle. (Adapted from ref. 42.)

## 3. Chemical Commonalities

The seven families of high temperature superconducting materials do not fall into a well-defined chemical group. However, one general feature that clearly distinguishes the high-$T_c$ set from the metals and alloys that dominated the low-$T_c$ era is apparent from the simple chemical sorting shown in Fig. 10. This follows the standard classification of the elements as metals or nonmetals. A plot of $T_c$ against the atomic fraction of nonmetal shows that all of the highest $T_c$ materials have at least 50% nonmetal content, whereas $Nb_3Ge$ and the alloys that dominated the low-$T_c$ era have <50% nonmetal. This plot follows the Edwards and Sienko classification of elements based on refractivity/volume ratios,[43] in which As and Ge are respectively a nonmetal and metal, but the above observation is unchanged if Ge is taken to be a nonmetal. In only one of the seven families, the iron arsenides, are some high-$T_c$ materials with <50% nonmetal content observed, and even here the highest-$T_c$ member of the family, $SmFeAsO_{1-x}F_x$, has 50% nonmetal content. In this context, the $AFe_2As_2$ family with $T_c$'s up to 38 K in $Ba_{0.6}K_{0.4}Fe_2As_2$ are notable as the only metal-rich materials to have exceeded the 23 K limit.

A well-known physical truism is that '*a good superconductor is a bad metal*', meaning that the electron-pairing interactions that give rise to superconductivity also diminish conductivity in the normal state above $T_c$. The equivalent chemical statement is thus that '*a good superconductor is mostly nonmetal*', meaning that the narrow bands and strong electron-electron correlations required for high temperature superconductivity are found in conducting materials with a high nonmetal content.

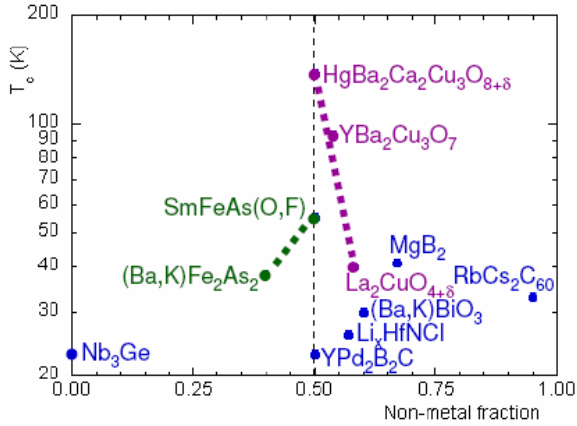

**Fig. 10** $T_c$ (on a log scale) plotted against the atomic proportion of nonmetallic elements for the seven distinct highest-$T_c$ materials plus $Nb_3Ge$, with other iron arsenide and cuprate types also shown.

As for many metal-nonmetal compounds, the electron distribution in high-$T_c$ superconductors and their parent materials can be represented to a first approximation by ionic formulae based on typical valence states, e.g. $(La^{3+})_2Cu^{2+}(O^{2-})_4$ and $Sm^{3+}Fe^{2+}As^{3-}O^{2-}$, although the distribution $Y^{3+}(Pd_2^{3+})(B_2C^{6-})$ implies some Pd-Pd bonding as is observed in the quaternary borocarbide crystal structure. Further consideration of the roles of the metal and nonmetal in the high-$T_c$ families shows that there are two limiting cases. The first is where only metal to nonmetal bonding is significant. These 'metal-nonmetal' families (cuprates, iron arsenides, bismuthates and nitride halides) have structures that follow simple ionic principles, with metal and nonmetal atoms bonded to each other but not to themselves, leading to near 50:50 metal:nonmetal compositions (between 40:60 and 60:40 for the high-$T_c$ materials on Fig. 10). The oxide and nitride halide superconductors have average anion charges of -2 and the presence of higher valent cations leads to nonmetal contents >50%. The $AFe_2As_2$ (and AFeAs) arsenides fall below the 50% limit because the magnitude of the anion charge is greater than that of the cations. From Fig. 10 it is apparent that the highest-$T_c$ members of the cuprates and iron arsenides are those with compositions closest to 50% nonmetal content, but the significance of this observation is unclear. An important chemical feature is that a specific metal-nonmetal pair is required to generate high-$T_c$'s in these families. No obvious analogues to the Cu-O and Bi-O combinations have been reported, but in the other two cases optimum pairings with $T_c > 23$ K/suboptimal pairings with $T_c < 23$ K are evident; Fe-As/Fe-P,Ni-P,etc. for iron pnictides; Hf-N/Zr-N,Ti-N for nitride halides. This demonstrates that superconductivity is very sensitive to the degree of nonmetal to metal charge transfer that occurs through orbital hybridisation. This group includes the most unconventional (least BCS-like) families, where metal-nonmetal orbital hybridisation is also important in strengthening the magnetic exchange that may mediate spin fluctuation superconductivity in cuprates and iron arsenides.

In the second 'nonmetal-bonded' limit, nonmetal to nonmetal bonding is important, and the cations play a subsidiary role. This is best exemplified by the $A_3C_{60}$ fullerides where the alkali metal cations act only as spacers for the fulleride anions and do not contribute to the electronic states near the Fermi level. $MgB_2$ is close to this limit although here some B to Mg charge transfer is important to the electronic structure. The presence of covalent bonding between nonmetal atoms leads to a lower anion charge per atom and hence higher nonmetal contents; $MgB_2$ contains $(B^-)_\infty$ sheets and the fullerides contain discrete $C_{60}^{3-}$ anions. This group of high temperature superconductors is more BCS-like, with nonmetal-nonmetal bonding leading to narrow bands and high vibrational frequencies, and hence strong electron-phonon coupling. However, the antiferromagnetism observed at the limit of superconductivity in fullerides may also be relevant.

The quaternary borocarbides are intermediate between the above two limits as they contain strongly bonded, discrete $B_2C$ groups but also show chemical specificity for Pd over Ni or other metals and have a 50:50 metal-nonmetal composition. Of the seven high temperature superconductor families, this is the only one to show prominent metal-metal bonding, and as the $T_c$ does not exceed that of $Nb_3Ge$, it could equally be regarded as 'low $T_c$'.

## 4. Future Prospects

The first 25 years of the high-$T_c$ era have been spectacularly productive, with a new $T_c \geq 23$ K superconductor family discovered every few years. As noted above, each family is chemically distinct from the others, so prediction of future discoveries is difficult, however some indicators are evident.

Based on the known high-$T_c$ materials, future high-$T_c$ superconductors are likely to have a substantial ($\geq 50\%$) nonmetal content. Very high $T_c$'s are associated with a specific pairing in the 'metal-nonmetal' group, and have nonmetal contents of 40-60% and structures in keeping with simple ionic bonding considerations. Neither element may have featured in previous high-$T_c$ families, as for Fe and As prior to 2008. Spin-spin fluctuations appear to be important to the superconducting mechanism in the highest-$T_c$ cuprates and iron arsenide materials where magnetic transition metals are needed. However, f-block magnetism may also be beneficial, as found in the intermetallic heavy fermion superconductor $PuCoGa_5$ which has $T_c = 18$ K,[44] so incorporation of nonmetals into such materials might prove a useful aid to future discoveries. Non-magnetic mechanisms based on other electronic instabilities may also emerge, as for the bismuthate superconductors based on suppression of charge disproportionation.

Chemical doping is required to suppress the spin or charge ordered ground state and induce superconductivity in the metal-nonmetal families. This is typically achieved through non-aliovalent substitutions or non-stoichiometry, but can sometimes occur through a fortuitous band overlap, as in $YBa_2Cu_3O_7$. However, disorder within the essential metal-nonmetal network tends to suppress superconductivity, and so additional parts of the structure that can be chemically tuned (the 'charge reservoir' in cuprates) are needed to obtain high-$T_c$'s. This is illustrated by the difference between the maximum $T_c$'s of 13 K for $BaBi_{1-x}Pb_xO_3$ doped at the essential Bi sites, and 30 K for $Ba_{1-x}K_xBiO_3$ doped at the secondary Ba sites.

The second group of high-$T_c$ materials is characterised by nonmetal-nonmetal bonding which may lead to very high nonmetal contents (potentially 100% for a purely organic superconductor). This group is limited to those elements that form strongly bonded covalent molecules and networks, typically B and C, although similar nonmetals such as N, O, Si, P, S could also be incorporated. Superconductivity is BCS-like in this group and the maximum observed $T_c$ to date of 41 K is in keeping with optimal BCS weak-coupling predictions. In these families the optimal electronic structures for superconductivity are achieved for the stoichiometric compositions $A_3C_{60}$ and $MgB_2$ (and also $YPd_2B_2C$) and chemical doping does not raise $T_c$.

Structurally, most of the high-$T_c$ families are based on layered arrangements, although the cubic structures of $A_3C_{60}$ and $Ba_{1-x}K_xBiO_3$ show that this is not a strict requirement. Two-dimensionality may be physically important for enhancing the pairing fluctuations needed for superconductivity. However, layered structures also offer far greater chemical and structural flexibility than three-dimensional networks, and so it is more likely that the optimal conditions for superconductivity can be realised. The highest-$T_c$ $HgBa_2Ca_2Cu_3O_{8+\delta}$ material represents one of approximately 30 chemically or structurally distinct cuprate subfamilies.

Predictions of the highest-possible $T_c$'s given the present known materials are extremely difficult. As estimates based on physical understanding of the mechanisms for high temperature superconductivity are still limited, a statistical approach may be the most realistic approach to gauge the likelihood of discovering new high-$T_c$ materials. Extreme value theory describes the probabilities of substantial deviations from the median in a large collection of values generated from a given set of rules, and is used to model the likelihood of occurrence of rare events such as mechanical failures or extreme weather.[45] Three limiting distributions are found, of which the Fréchet type is appropriate to a variable bounded by a lower limit such as $T_c > 0$ here. The Fréchet probability distribution function is:

$$f(T_c) = (\alpha/\beta)(\beta/T_c)^{1+\alpha}exp(-(\beta/T_c)^\alpha),$$

and the cumulative distribution function is:

$$F(T_c \leq T) = exp(-(\beta/T)^\alpha).$$

Fig. 11 shows the distribution of maximum $T_c$ values from the seven families known to have maximum $T_c \geq 23$ K (as shown in Table 1) plus $Nb_3Ge$. Assuming that the total number of presently known superconductor families is $N_{Tot} \sim 1000$, $f(T_c)$ values were obtained in 10 K intervals as $n/N_{Tot}$ where n is the number of superconductor families with maximum $T_c$ in the ranges $23 \leq T_c/K < 33$, $33 \leq T_c/K < 43$, etc. The $f(T_c)$ fit shown has parameters $\alpha = 1.5$ and $\beta = 5$. Improved estimates of these parameters could be obtained by analysing maximum $T_c$ values in the low-$T_c$ region.

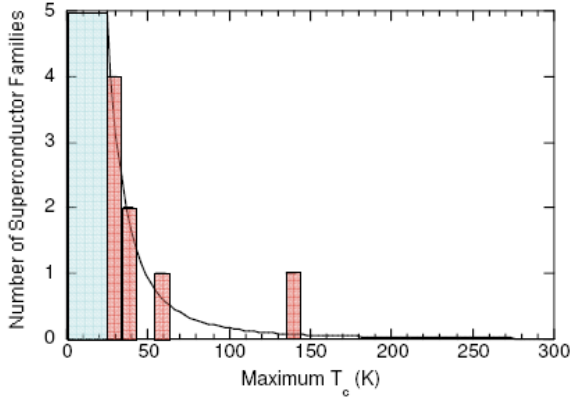

**Fig. 11** The distribution of superconductor families according to their maximum $T_c$, binned in 10 K intervals for $T_c \geq 23$ K materials. The curve has equation $1000f(T_c)$ with Fréchet function parameters as shown in the text.

The probability of a superconductor family having maximum $T_c > T$ is $(1 - F(T_c \leq T))$ so from the above parameter values, the probability of a newly discovered superconductor family having maximum $T_c > 100$ K is 1%, decreasing to 0.2% for $T_c > 300$ K (room temperature superconductivity). These statistical estimates appear up to an order of magnitude too high as only one presently known family has $T_c > 100$ K, so ranges of 0.1-1% and 0.02-0.2% may be more realistic, and even these should be treated with appropriate caution. On a statistical basis, several hundred new families of superconductor may have to be discovered to find another $T_c > 100$ K material, and perhaps many hundreds or a few thousand to realise the possibility of superconductivity at room temperature. However, more targeted approaches based on knowledge of known materials as presented here, or perhaps on new theoretical insights, should reduce the chemical space to be explored.

**Conclusions**

Seven distinct superconductor families with ambient pressure $T_c$'s that equal or surpass the historic limit of $T_c = 23$ K in $Nb_3Ge$ have been discovered in the last 25 years. These high-$T_c$ families are all nonmetal-rich unlike the metals and alloys that dominated the earlier low-$T_c$ era. The high-$T_c$ superconductors are chemically diverse, but broadly fall into 'metal-nonmetal' and 'nonmetal-bonded' groups. Materials in the former group are based on an essential metal-nonmetal pair e.g. Cu-O, Fe-As and approximate to an ionic description where only metal-nonmetal bonding is significant and the metal:nonmetal ratio is near 50:50. Superconductivity is optimised in non-stoichiometric materials through chemical doping and leads to the highest known $T_c$'s in the cuprates and iron arsenides, probably through a magnetic spin fluctuations mechanism, although other coupling mechanisms such as charge fluctuation interactions may also operate in this group. 'Nonmetal-bonded' materials contain covalently bonded molecular or extended anion networks and are BCS-like superconductors with $T_c$'s up to ~40 K.

Replicating the principal features of the 'metal-nonmetal' group may provide the best chemical guidance to the discovery of future high-temperature superconductors at present. Layered structures are advantageous for introducing strong pairing fluctuations and provide a good opportunity for chemical and structural optimisation of $T_c$. The distribution of $T_c$ values for the known high-$T_c$ families follows a typical extreme values distribution and a statistical analysis suggests that the probability of a newly discovered superconductor family having maximum $T_c$ > 100 K is ~0.1-1%, decreasing to ~0.02-0.2% for room temperature superconductivity.

**Acknowledgements**

I acknowledge support from EaStCHEM, EPSRC and the Leverhulme Trust, and the use of the Chemical Database Service at Daresbury.